# The Cycle of Value

## A Conservationist Approach to Economics

**Author:** Nick Harkiolakis, New England College

**Abstract**

A representation of economic activity in the form of a law of conservation of value is presented based on the definition of value as potential to act in an environment. This allows the encapsulation of the term as a conserved quantity throughout transactions. Marginal value and speed of marginal value are defined as derivatives of value and marginal value, respectively. Traditional economic statements are represented here as cycles of value where value is conserved. Producer-consumer dyads, shortage and surplus, as well as the role of the value in representing the market and the economy are explored. The role of the government in the economy is also explained through the cycles of value the government is involved in. Traditional economic statements and assumptions produce existing hypotheses as outcomes of the law of conservation of value.

# 1. Introduction

One of the most challenging concepts in economics since the first accounts of economic thought is that of *value* (Benveniste & Scheinkman, 1979; Brown, 1984; Kallis, Gómez-Baggethun & Zografos, 2013, Lakdawalla et al., 2018; Patterson, 1998; Smith, 1976). Defining it is difficult primarily due to its subjective nature, at least as perceived by individuals, organizations, and societies. Individuals have different perceptions of the value of a piece of art, friendship, chocolate, clean air, democracy, relaxing by the beach, owning a yacht, watching a good movie, being healthy, life after death, etc.

Value can be seen as a measure of appreciation for something, a monetary worth, benefit, usefulness, etc., relative to something that is not presumably at its desired level or missing completely (Bowma, & Ambrosini, 2000; Brandenburger & Stuart, 1996; Smith, 1976). In this sense value is more like beauty which is in the eye of the beholder since it requires an outside observer for its existence. This subjective observer could for all purposes be any living entity or groups of entities that depend on their environment for survival and growth.

Written accounts debating the definition of value can be traced as far back as the fourth century BC when Aristotle distinguished between value of "use" and value of "exchange" (Johnson, 1939). One could value their clothing because of their function/use (keeping them warm and protected from the elements) but also could exchange them if needed for food or services with another who might be interested in having them. After this initial treatment, it took up until medieval times (13$^{th}$ AD) for Aquinas (Grassl, 2010) and Scotus (Hare, 2000) to give *value* an ethical flavor as something one deserves and can receive by producing work that could result in a product or service.

From the 17$^{th}$ century onwards, a race to define value is taking place in attempts to equate it to utility, labor, land, market price, and combinations and subdivisions of these (Bellofiore, 1989; Cheshire & Sheppard, 2017; Foley, 2000; Vaughn, 1978). Along the way the dynamic influence of time became evident and concepts like supply and demand, and marginal utility surfaced in an attempt to explain the behavior of businesses, markets, and economies at large (O'Donoghue & Somerville, 2018; Zhao et al., 2019). While the reductionist approach in defining value produced a lot of breakthroughs, the economic challenges we face today are an indication that we haven't really managed to develop a theory of value, at least with the success other concepts have received in hard sciences like physics, chemistry, and mathematics (Fukumoto & Bozeman, 2019; Shogren & Taylor, 2020).

Value is an abstract construct that has no physical manifestation. It does not exist in the physical world, at least as an observable physical entity, and for that reason it cannot be sensed or detected through our senses or instruments. Nevertheless, its existence in language makes it an information-carrying element that is ordinal in nature, meaning we can order it along a continuum from low to high (Bergemann et al., 2020; Fenwick et al., 2020; Viet et al., 2018). As such it is suitable for comparisons and, as we will see later, for arithmetic manipulation.

A contradiction between the assumptions of traditional economics and the realities of the market results in much of the inequalities and inefficiencies we observe in today's societies. The core contradiction is the assumption of equal levels of rationality among individual agents and their competition as the judge of market direction (Arrow, 1990; Rabin, 1990). If everyone is equal (even in terms of rationality), how can it be possible to have some who are winners and some who are losers? The answer is in every other factor that plays a role in what happens, including the operational environment and resources available (Foley, 2004). Even this consideration is not going to work, so the primary assumption about equal rationality is a false premise that does not reflect reality.

Another assumption in traditional economics is that market outcomes are fair/equitable (Falk et al., 2003; Jost et al., 2003). The increasing divide between the poor and the rich would beg to differ. Finally, the notion of "free choice" is prevalent in traditional economics (Altman, 2006; Gronskas, 2003). Social unrests and inequalities again will beg to differ on how "free" someone's choice is given the constraints imposed in many cases. How free is someone to choose the education they want when they are poor?

## 2. The Conservation of Value

Consider the simplistic case of a meadow where sheep run wild and graze the grass in a random fashion and as they please (top of Fig. 1). Then humans come and start corralling the sheep into an enclosed space. The question that concerns economics is why they did so when they could just as well milk, shear, and eat them where they were. The answer is because it is more valuable/convenient to have the sheep corralled where you can easily access and exploit them for their milk, wool, and meat. We should note here that it was the same number of sheep in the field and in the corral but for us humans the latter arrangement is more valuable than the former.

Looking at what we described in more detail, we can say that there was something of *value* out there (in nature), we exerted some effort to collect and organize it, and we ended up with something that is more valuable to us than before. Granted we spend some energy to go and collect the sheep, but we can easily envision the beneficial possibilities of our new situation. We also might have used technology (rope, shoes, etc.) to run in the field and chase the sheep (we could even use a trained dog). With some effort again and using different tools, we can easily convert what we have in produce (milk, wool, meat) and make some money by selling it. Fig. 1 depicts the process we just described by representing the labor/capital we spend with the dollar sign and the energy/effort we lost in the process with the explosion graphic. Our goal through the whole process (excluding other factors) is to hopefully end up with more value (money in Fig. 1) than we started.

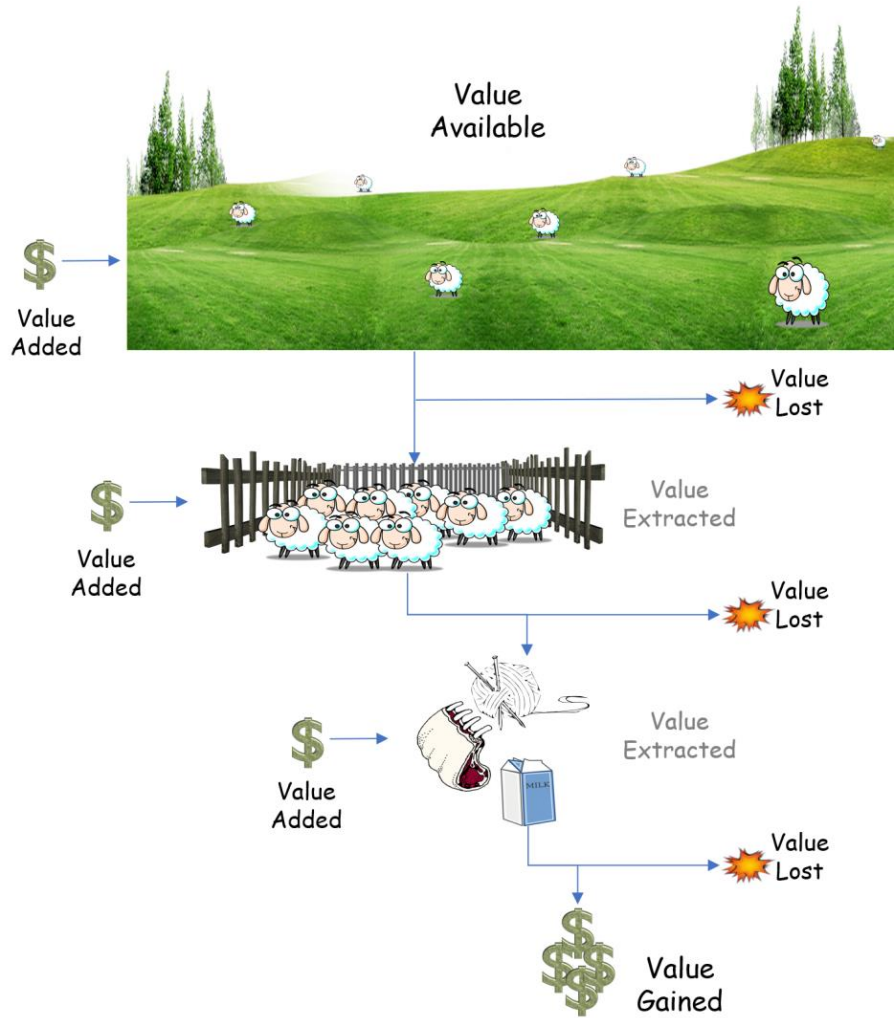

Fig. 1. The value creation process

In economics as in any other science, it helps to abstract the phenomena we observe as much as possible, to simplify their complexity to a level that is easy to express in formulas and develop theories to describe them. Abstractions also allow us to generalize our descriptions and findings to wider domains. It is easy to imagine that what we described with sheep can be easily applied to other livestock (fish, birds, etc.) and natural resources. In that respect, Fig. 1 can be seen as a special case of the more general process depicted in Fig. 2 where the various components of the process are represented as tanks of *value*. The choice of the word *value* here was made to emphasize the anthropogenic approach we are following in studying what is called the production process. The subjective connotations of the word will be addressed later when the consumer and producer perspectives are presented.

Value is seen here as representative of a process within the physical and social world. In this respect, value is not something that exists in an instance of time independently of its past or future but rather the culmination of an agent's activity when interacting with the natural and social environment. In other words, it is a potential to act in a situation. As a working definition, value can be seen as *the potential to act in an environment*.

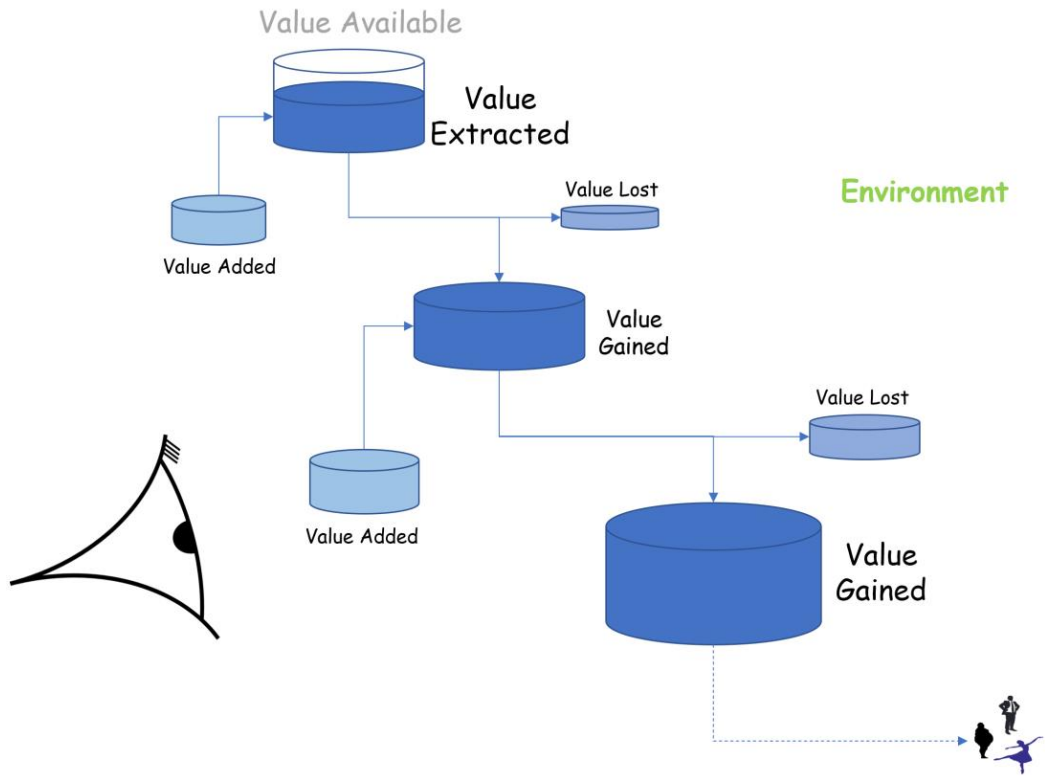

Fig. 2. The conceptualized value creation process

While the extracted and gained values will be seen as static/state elements, the added and lost values will be seen more as dynamic/flowing elements (changes of static elements). This conjecture will become clearer later when we compare the conservation of value with the law of conservation of energy. As Value Added, we consider everything that went into the process, including money (seen here as proxy for inherited, borrowed, and accumulated value from past transactions), labor, and equity of any other form. As Value Lost, we consider anything that cannot be returned to the cycle, including energy (mainly heat), money (transaction costs, debt), waste, etc.

We can easily deduce that the whole process can be seen as a balance sheet between Extracted and Gained Value through exchanges of Value Added and Value Lost. Fig. 3 depicts what will be called *the law of the conservation of value*. The value that exists somewhere (for example, in nature) requires an amount of *Value Added* (VA), that will be used to access *Value Extracted* (VE). In the process, some will dissipate in the environment as *Value Lost* (VL) and we will end up with a net amount of *Value Gained* (VG). This statement can be expressed as:

"Value Added" + "Value Extracted" = "Value Lost" + "Value Gained"

and in acronym form:

$$VA + VE = VL + VG \qquad (1)$$

What the equation describes is the transformation of value. A visual representation of the law is depicted in Fig. 3 where the equation indicates the balance between the two sides of the equation. The balancing of values can be seen taking place at the various stages of the transformation/production process and can be even applied additively across multiple steps/cycles of the process.

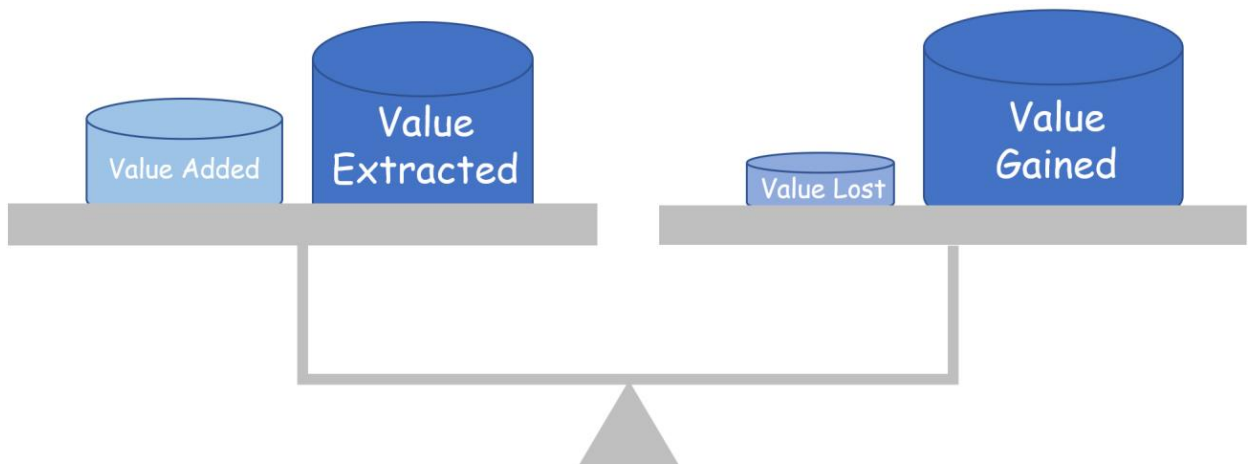

Fig. 3. The law of conservation of value

The continuous process in Fig. 2 eventually ends (hopefully) with the distribution of more value to society or at least parts of it. For some, the outcome will be material for their immediate needs in the form of food, clothing, entertainment, etc., while for others it will be in the form of money/profit. The latter is interesting because it is not value in the typical sense of present value but rather credit for value they can claim in the future. This idea of value reserves that can be used in the future is one of the cornerstones of modern economics and a distinct characteristic of human societies (Foss, 1997; Lewin & Phelan, 2000; Maxfield, 2008).

Considering the notion of crossing past, present, and future in terms of value flow, one can borrow value from the future (like by borrowing from a bank), do their production magic in the present, and when the future becomes present return it with interest (Fig. 4). Danger lurks in the unpredictable nature of the future, which brings uncertainties with respect to our ability to effectively break even the balance sheet of values.

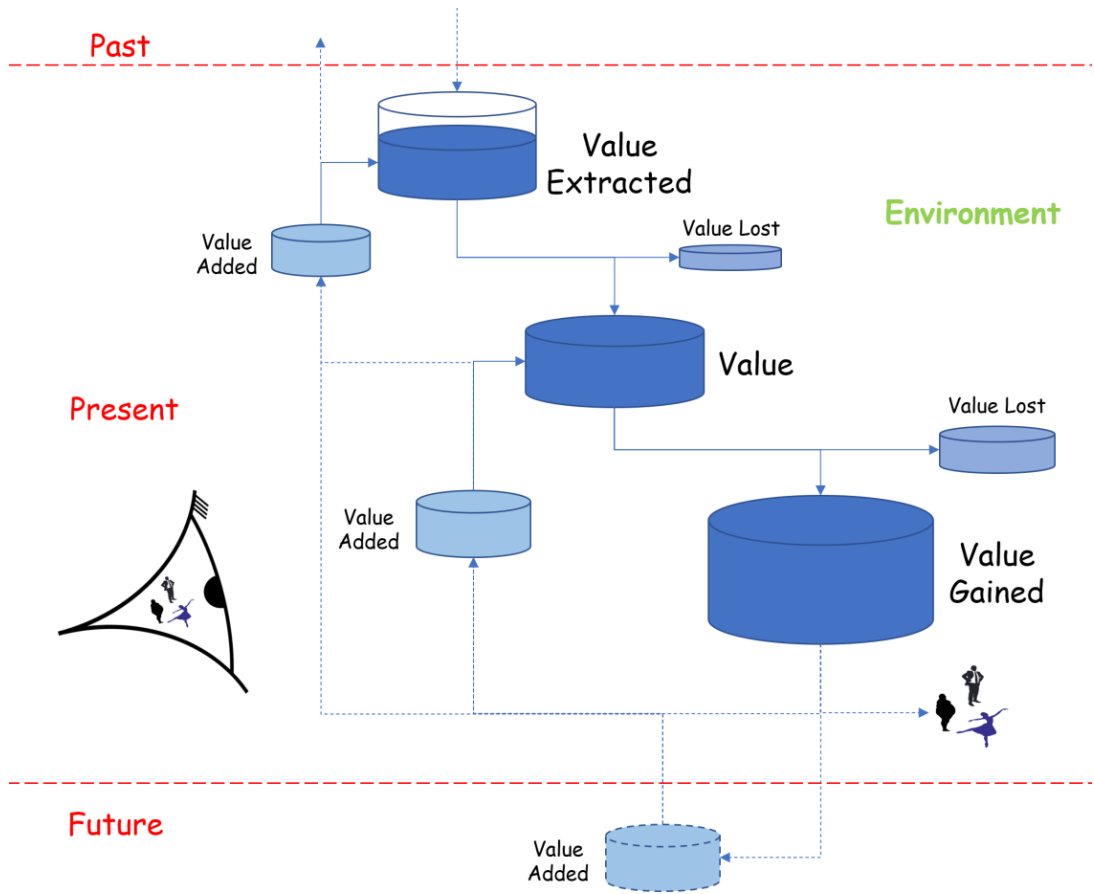

Fig. 4. The value creation process over time

The reason the word "value" was preferred over "utility" is simply to avoid the association of utility with utilitarianism with its morality connotations that would assume an intention to fair and just distribution of utility/wealth. This contradicts the free-market realities that are better reflected as the survival of the fittest (Park & Lee, 2021). With respect to *marginal utility* (adding satisfaction/usefulness), the assumption of seeking to marginalize (maximize) something is suggestive of our intention to increase it by adding something extra. This direction will make the notion of equilibrium an antagonistic battlefield where opponents reach a stalemate, and no one can gain more than the other despite their continuous efforts. In this paper we will look at the economic situation as a constant flow of value where its conservation is established as "law". In other words, we see our existence with nature as a closed system where its total value is preserved while internally it can be distributed differently. Looking at Earth as a spaceship in the vacuum of the cosmos is probably the best analogy of what is considered in this paper a closed system of value or a system with a finite amount of value.

Considering Fig. 2, a cyclic process of the flow and conversion of value can be visualized. VA and VE work together to produce VG and VL. The produced VG then can be consumed or used as VA for another cycle of conversion/production. Fig. 5 depicts what will be called from now on the *cycle of value*. This analogy will be used in future sections to explain the economic perspectives of governments, organizations, and even individuals as they engage in economic activities. While the analogy to a great extent follows the systems theory concepts of stocks and flows, it will be shown here how it differs from them significantly (Baghaei Lakeh & Ghaffarzadegan, 2016; Fischer & Gonzalez, 2016).

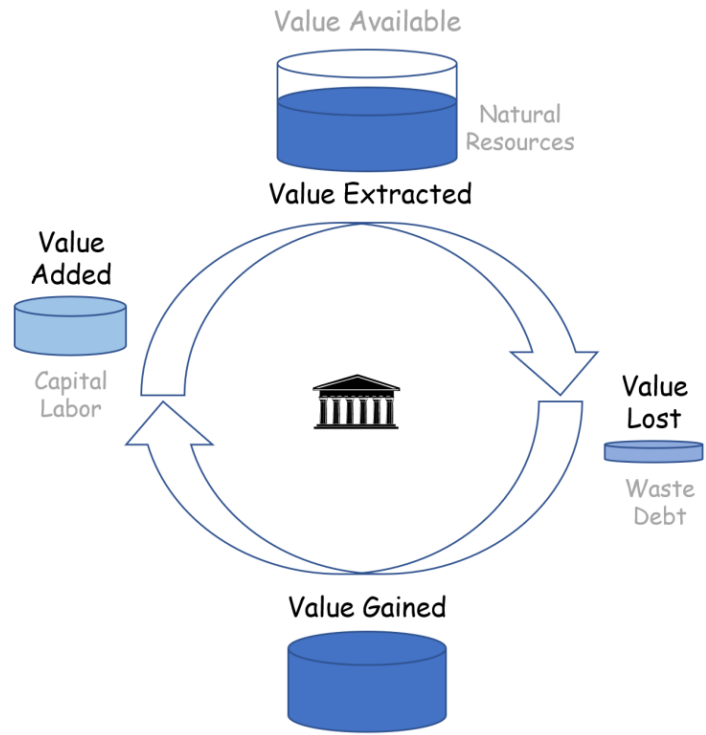

Fig. 5. The cycle of value

The cycle of value is an alternative visualization of the law of conservation of value (Fig. 3) and it will be used from now on as a more appealing visualization of the law due to its simplicity and the inherent notion that value creation is a recurring process. Something that should not be confused with is that the representation of Fig. 5 is not a stepwise sequence in time in which we first add the value then we extract value, then we lose some, and we eventually end up with some. The cycle is dynamic in that we add and lose value continually from beginning to end. Fig. 5 and equation (1) describe the accounting of the elements, similar to how profits and losses in a balance sheet are accumulated figures for gains and losses throughout a year. The time element is not apparent, but we will soon see how it can be introduced.

Having a cycle that leads to more cycles and when multiple entities are accessing the VE could intuitively suggest that some coordination and control will be required so things do not get out of hand. This is where government and political institutions with regulations and control mechanisms come into the picture (center of Fig. 5). Government's role is to ensure this cyclical transformation keeps moving in the right direction (clockwise in our case) so that citizens can continue and prosper in an ever-increasing way. In return, government will take a share of the VG for its future function and leave the remaining to the individuals and organizations that invested in the process.

**3. Value and Energy**

From a physical perspective one can see the analogy of the cycle of value to a form of energy conservation. As Fig. 6 depicts, we can assume that the VE and the VG are static/state characteristics in the sense that they represent something that exists at a certain instant of time, suggesting, by analogy to physics, they play the role of potential energy (EP). On the other hand, VA and VL represent value flow through time and as such can be analogous to kinetic energy (EK). When an object in physics moves from one position to another in a conservative field of forces, we can say that according to the law of conservation of energy:

$EK_1 + EP_1 = EK_2 + EP_2$

By analogy (Fig. 6), we get:

$VA + VE = VG + VL$

which is what we have already established by equation (1).

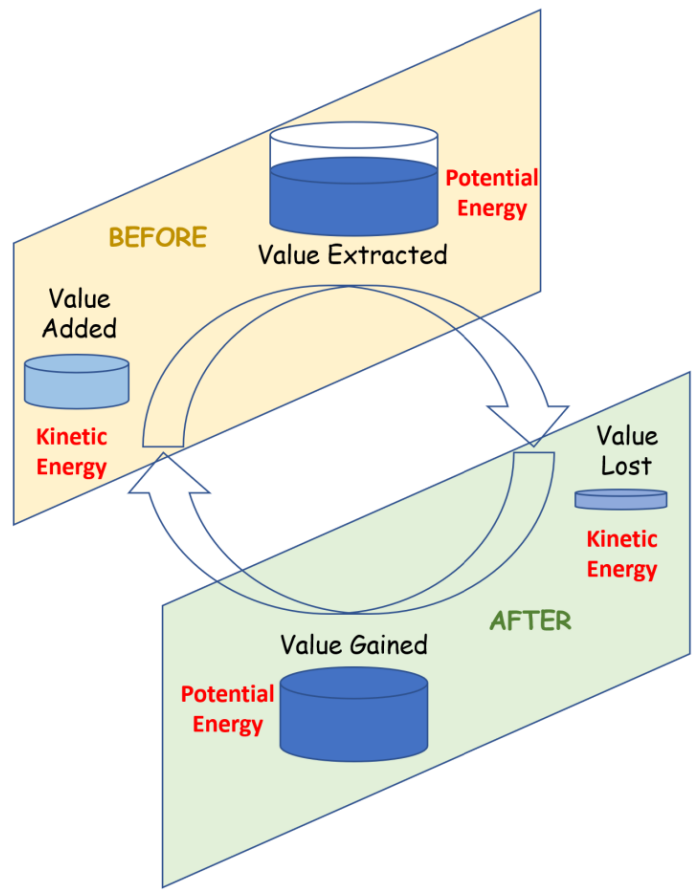

Fig. 6. The conservation of value as energy

A necessary clarification is required here in that the VG and VE are not exactly state quantities but rather differences of state quantities. For the whole process to start we must have had some initial amount of value (Value Initial) available to us as potential/labor/capital/etc. From that we diverted some as VA to the transformation process and eventually ended up with a final amount (Value Final) of value. So:

Value Gained = Value Final – Value Initial

Similarly, the VE is the difference from what was initially available (for example, in nature) minus what we leave after the extraction. These "differences" are analogous to the potential energy in physics as we always measure potential with respect to a reference point. For example, when it comes to the gravitational potential, we sometimes use the surface of the Earth as our zero point/level while other times we consider its center.

### 4. Marginal Value

Considering value as the variable of interest, the velocity (speed really as there value is considered a scalar quantity) of value is the rate of change of value over time. This quantity will be called here marginal value in correspondence to the marginal utility in economics (Kauder, 2015). The difference between the two is that marginal value represents rate of change in time while marginal utility represents rate of change per quantity produced (Kauder, 2015; Li & Hsee, 2021; Veblen, 1909) or the increase in utility when an additional unit of output is achieved. With a steady flow of production, as standard economics assumes, there will be a direct correspondence between the unit of production and the time it takes to produce it, making marginal value similar to marginal utility when added consumer satisfaction is considered. It will be seen here that considering time as independent variable offers advantages in terms of expressiveness in economic terms.

Before moving further, it is worth mentioning here that the concepts of marginal utility and "diminishing marginal utility" that will be seen later are not based on any mathematical formulation and proof that is confirmed by a hypothesis testing process (Benabbou et al., 2019a; Benabbou et al., 2019b). They are propositions that are true by the nature of humans engaging in voluntary or coerced exchange where they try to maximize their utility. The assumption is made that rationally acting individuals always try to maximize satisfaction/utility (Foley, 2004; Simon, 1979). In other words, between two satisfying alternatives one will always chose the one that provides more satisfaction.

For the following discussion the Lagrange notation will be used to indicate the derivative of a variable ($x´$) instead of the Leibniz ($dx/dt$) form. Let's consider first that for a short time interval or unit of time ($dt$ tends to zero) a Marginal Value Added (VA´) has been injected into an existing cycle that accessed a Marginal Value Extracted (VE´). In the process a Marginal Value Lost (VL´) is lost. As a result, we will end up with Marginal Value Gained (VG´) (Fig. 7). By differentiating (1) we get:

(VA + VE)´ = (VL + VG)´

or

VA´ + VE´ = VL´ + VG´

or

VG´ = VA´ + VE´ – VL´

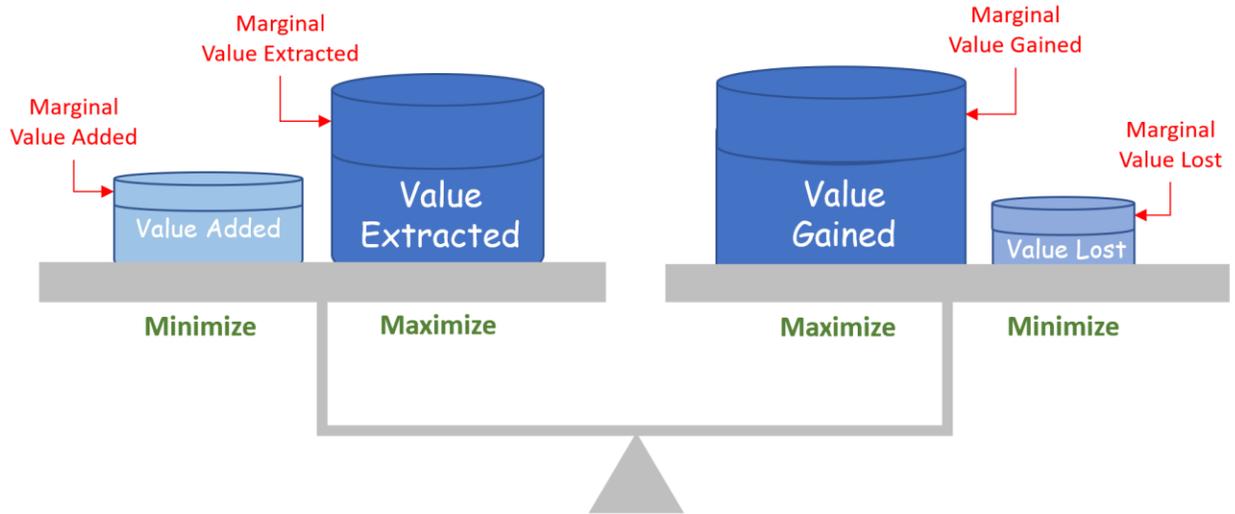

Fig. 7. The balance of marginal values

The goal in economics would be to maximize gains and minimize losses. If an abundant source of extracted value is assumed, like nature according to neoclassical economics, then the rate of the extracted value will be zero simply because the amount of value in nature will not change (Daoud, 2018; Reda, 2018). This assumption is not farfetched at least when one considers small time horizons like days, weeks, and even months.

The last equation then becomes:

VG′ = VA′ – VL′     (2)

This means that the Marginal Value Gained will be equal to the Marginal Value Added minus the Marginal Value Lost. Considering that the goal of agents is to maximize their Value Gained, then VG′ = 0 when the maximum of Value Gained is reached. In this case, (2) becomes:

VA′ = VL′     (3)

The maximization of Marginal Value Gained (faster rate of gaining value) can be achieved when Marginal Value Added is equal to Marginal Value Lost. We will see the consequences of this result later when it will also become evident that considering time rates instead of production unit rates, which economics is using now, results in more naturally interpretable results.

The law of conservation of value for one cycle can also be applied to multiple ones that run simultaneously by simply adding the various components. Equation (2) will then suggest that the sum of the Marginal Value Gained through all cycles will be equal to the sum of the Marginal Value Added minus the sum of Marginal Value Lost. Similarly, (3) will suggest that to maximize our gains we need to make sure the sum of the Marginal Value Added is equal to the sum of Marginal Value Lost.

## 5. Rate of Change of Marginal Value

Having considered the speed/rate of change in time (marginal value) in the cycle of value, the next step would be to consider the rate of change of marginal value. The equivalent concept in terms of utility is that of diminishing marginal utility, which is defined as the rate of decline of marginal utility. In the case of the cycle of value, the term Speed of Marginal Value (acceleration of value) will be introduced. Alternatively, the term *Diminishing Marginal Value* can be used when the rate of change of marginal value is negative and *Increasing Marginal Value* (value flow decelerates) when the rate of change of marginal value is positive (value flow accelerates). The latter case is also of interest as it could reveal missed opportunities for faster rates of growth. Considering the law of conservation of value in terms of the speed of change of marginal value, we will have that the Speed of Marginal Value Added (VA´´) affects the Speed of Marginal Value Extracted (VE´´) and as a result it will impact the Speed of Marginal Value Gained (VG´´) and the Speed of Marginal Value Lost (VL´´).

This can be expressed as:

$$VA´´ + VE´´ = VL´´ + VG´´$$

or

$$VG´´ = VA´´ + VE´´ - VL´´ \quad (4)$$

If we were to consider increases, then (4) can be interpreted as the Increasing Marginal Value Gained being equal to the Increasing Marginal Value Added plus the Increasing Marginal Value Extracted minus the Increasing Marginal Value Lost. For decreases of speed, Diminishing instead of Increasing can be used.

If now we assume the neoclassical view of an abundant nature, similar to (2), we will get:

$$VG´´ = VA´´ - VL´´ \quad (5)$$

In the case where Marginal Value Gained reaches its maximum, we will have:

$$VA´´ = VL´´ \quad (6)$$

The maximum or minimum of Marginal Value Gained will happen when the Speed of Marginal Value Added is equal to the Speed of Marginal Value Lost.

The second derivative (acceleration) with respect to quantity is where traditional economic analysis usually stops. Considering the third derivative has not become mainstream yet. In physics though, the situation is different as the third derivative is typically seen as a *jolt* or *jerk*. Since there is no "marginal jolt" in economics, the *jolt of value* will be defined here as the rate of change of the speed of marginal value. This is like a kick/jumpstart in the cycle of value. It could be a sudden infusion/injection of value in the cycle similar to money infusions that governments can make to jumpstart their economies. Examples include stimulus checks to consumers or sudden taxation when the government wants to remove value from the economy.

Equation (4), in terms of Jolt Value will become:

$$VG´´´ = VA´´´ + VE´´´ - VL´´´$$

with similar forms for (5) and (6).

Before delving more into the usefulness and application of variables discussed up to now in the cycle of value, it is worth having a visual appreciation of how these terms relate and how they can be interpreted when a certain equation for the expression of value is considered. Value up to now has been seen as a generic term that represents something that can be exchanged and stored. Fig. 8 display the speed (V´), acceleration (V´´) and jolt (V´´´) of value in terms of their sign and the interpretation that it implies.

| | SIGN | | | | | | | |
|---|---|---|---|---|---|---|---|---|
| V´= ValueSpeed | + | + | + | + | - | - | - | - |
| V´´=ValueAccelerat. | + | + | - | - | + | + | - | - |
| V´´´= ValueJolt | + | - | + | - | + | - | + | - |

| | INTERPRETATION | | | | | | | |
|---|---|---|---|---|---|---|---|---|
| V´= ValueSpeed | The speed of change is positive | | | | The speed of change is negative | | | |
| V´´=ValueAccelerat. | and increasing | | and decreasing | | and increasing | | and decreasing | |
| V´´´= ValueJolt | fast | slowly | fast | slowly | fast | slowly | fast | slowly |

Fig. 8. Value motion variables

In closing this section, it is worth pointing out that the rates of change which refer to differentiation have their equivalent "inverse" as integration. In calculation total value during a period of time where multiple runs of the cycle of value take place, we can use integration of the formula that is adopted for value. This discussion is left for a future publication.

## 6. The Market

With respect to the cycle of value in a market perspective, multiple entities/agents are considered as an interacting closed group (Fig. 9). In the case of multiple actors in a market environment, the cycle of value will take the form (the index M is for market):

$VA_M + VE_M = VL_M + VG_M$

Each term in this equation is the sum of the corresponding terms of the participating entities. If we are interested in marginal values, then by differentiating the previous we get:

$VA´_M + VE´_M = VL´_M + VG´_M$

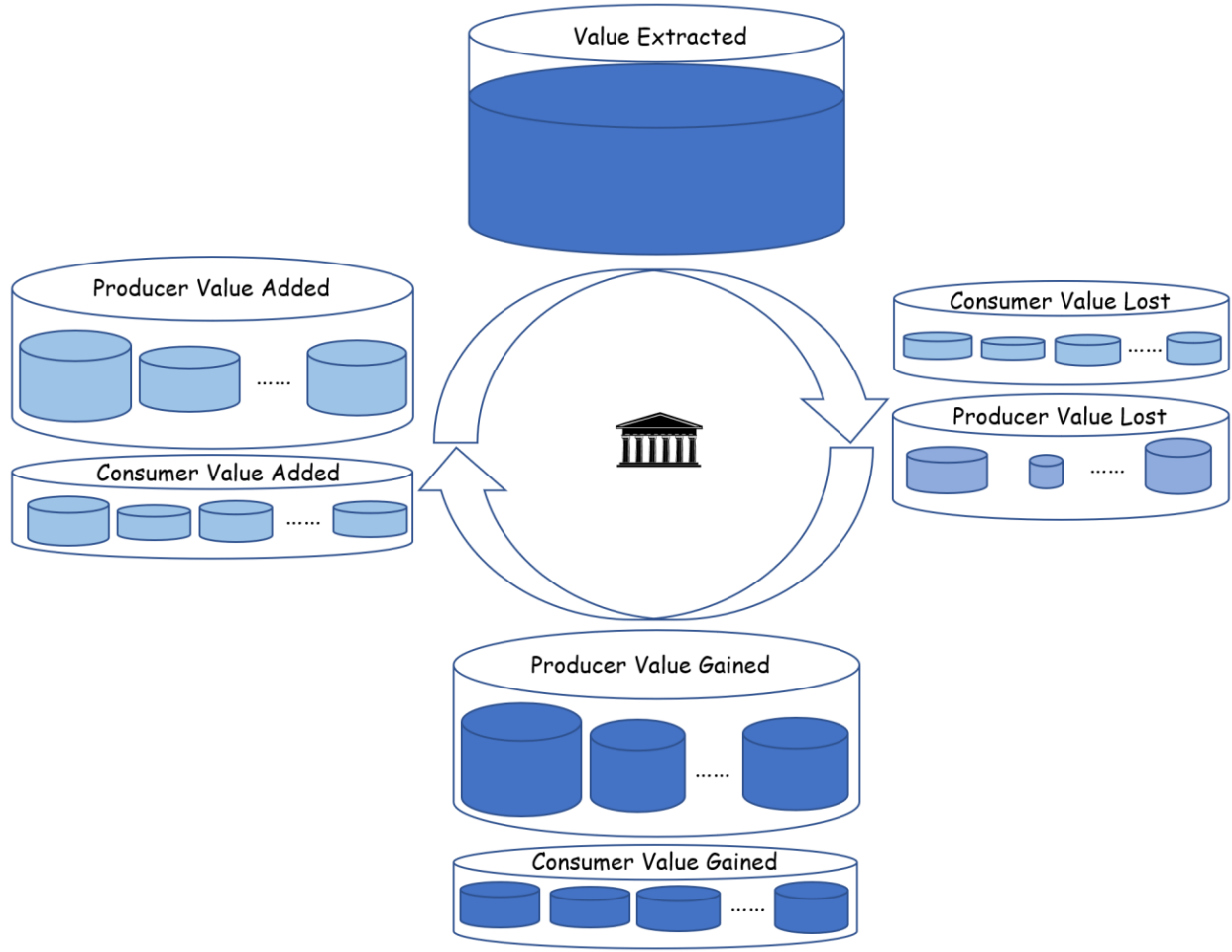

Fig. 9. Market and Economy

Assuming an abundant source of value for extraction, as it is typically assumed of nature ($VE'_M = 0$), the previous equation becomes:

$$VG'_M = VA'_M - VL'_M$$

This result suggests that the Marginal Value Gained in the market (market growth rate) is a function of the Marginal Value Added and Marginal Value Lost. If we want the market to grow, then we need to add value faster than we lose it.

Another interesting result can be obtained if we consider a fixed market size (the numbers of consumers and producers is constant - logical for short time periods) and a certain stability, then we can also assume that the Value Gained does not change or it can reach a maximum ($VG$ = max, when $VG'_M = 0$). At that point:

$$VA'_M = VL'_M$$

For a stable market we need to match the speed at which we lose value with the speed we add value.

Balancing the cycles of value of the various transactions to adhere to the law of conservation of value can be challenging due to the multiplicity of the entities in a market. It becomes difficult to realistically model agents with different experiences, priorities, and intentions. Representing reality is very difficult as we rarely know exactly what the VE is, how much we need to add to get it, how much we will lose in the process, and how much we will eventually gain. In other words, there is an uncertainty in every term of equation (1) which we will represent here as error terms. We will have error terms for Value Added (VAerror), Value Extracted (VEerror), Value Lost (VLerror), and Value Gained (VGerror). The error quantities here are assumed to have the error sign included in them.

The law of conservation of value for one cycle in its more accurate mathematical expression will become:

VA + VAerror + VE + VEerror = VG + VGerror + VL + VLerror

It can be seen from the above that there are multiple sources of error that should be accounted for. While this is true for the single case cycle, it is less of an issue for the market case where multiple entities participate in the same cycle. This is because the law of large populations will ensure that we have as many deviations up as we have down and in this way the sum of the error terms will tend to diminish. In statistical terminology we can say that as long as the sphericity is low, and the deviations from the mean are random (do not follow a pattern), equation (1) holds as representative of multiple cycles of value like a whole market.

## 7. The Economy

Expanding the cycle of value to the economy at large, each entity/agent in the economy can be considered going through its own cycle of value at each moment in time while interacting with each other and with nature (Fig. 10). Nature can be considered a passive agent in this scenario that only provides value (at least as long as it lasts). An agent X might want to access the VE value of another agent Y or nature. That could happen when, for example, X pays Y ( money if a form of VA) to do some work (VE), the result of which is VG for X. In the process, some value will be lost to the environment (VL) either as waste, taxes, insurance payments, or something else. The situation could involve organizations or even whole countries when they interact with other organizations or countries. The government and/or international monitoring institutions play the role of coordinator/controller in this situation.

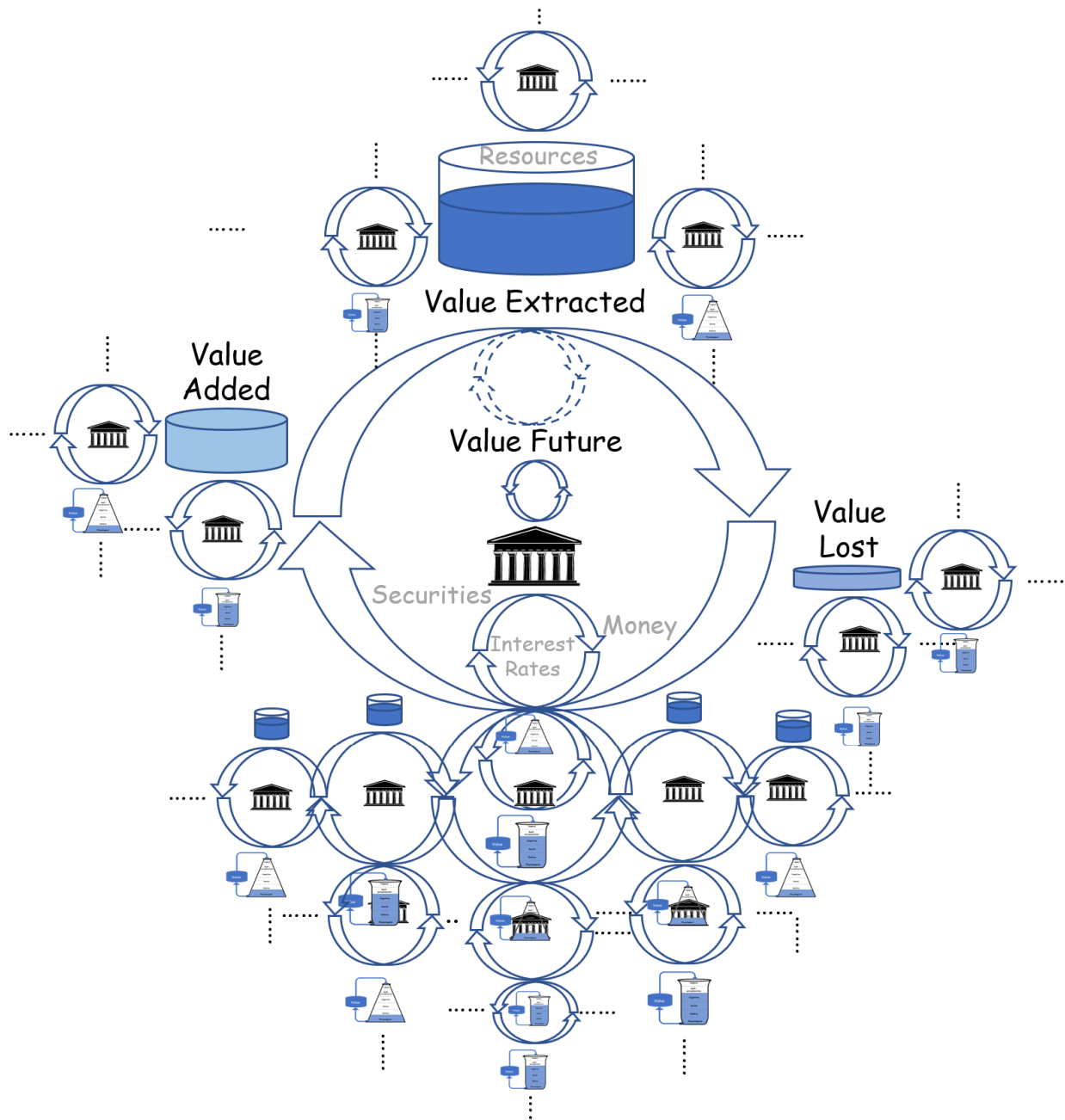

Fig. 10. The economy at the level of a country

Additionally, there could be multiple agents trying to access the same source of value and either compete or cooperate. There could be agents that consider some of the VL of other agents as their VG. The lost value of one agent could be in the form of debt they pay to another agent, for example, a bank. In this case, the bank/agent gains some of the VL of the agent who might have borrowed money before. Financial institutions can reliably be described in cycles of value. A bank could borrow money from a central bank as VG for its purposes. It could then lend that money (VG) to other banks, organizations, and individuals. In this case, the VG of the lending bank will play the role of VE for the borrowing entities. In return, the bank gains value from the lending process through the interest it charges, while in the process it loses some value as it has to pay for its facilities, employees, etc. and also return some money with interest to the central bank.

## 8. Government as Controller

A core ingredient of the cycle of value for the economy is the control mechanisms that are needed to ensure its continuous flow. Here is where government plays its critical role. Inefficiencies in the cycle (like economic downtimes) can be addressed by decreasing interest rates at the federal government level, by printing money, or by establishing laws and regulations to support economic growth. The question is what is the government's source of value that it would "generously" allow organizations and individuals to access. In the past, governments would accumulate gold whose value was perceived as timeless. It would then print money as a proxy to gold and allow its constituents access to them in the form of loans. These loans would be directly available to the banking system that served as intermediary between the government and organizations or citizens. The money would be invested to produce more wealth than what was borrowed, allowing organizations and individuals to repay the banks and keep the remaining as profit.

The banks would then pay back the government the money they borrowed plus interest and keep the remaining as their profit. As a result of economic activity, the government would increase its money deposits through taxes after subtracting the costs of sustaining its operations (Fig. 11). This extra amount of money could then be invested in services to its citizens and maybe purchasing more gold. The process is straightforward enough to the point of the United States (US) government realizing that they really did not need the gold in the cycle, so they just decided to do away with it altogether. From that point on, the US government just printed money any time it was needed to stimulate the economy and withdrew money to slow it down. The rationale for eliminating gold from the cycle was the sheer faith that the US brand was enough to ensure the cycle will keep going for ever. Money will be printed at some point in time and be returned at a future point in time. In this way, the "future" becomes the source of value.

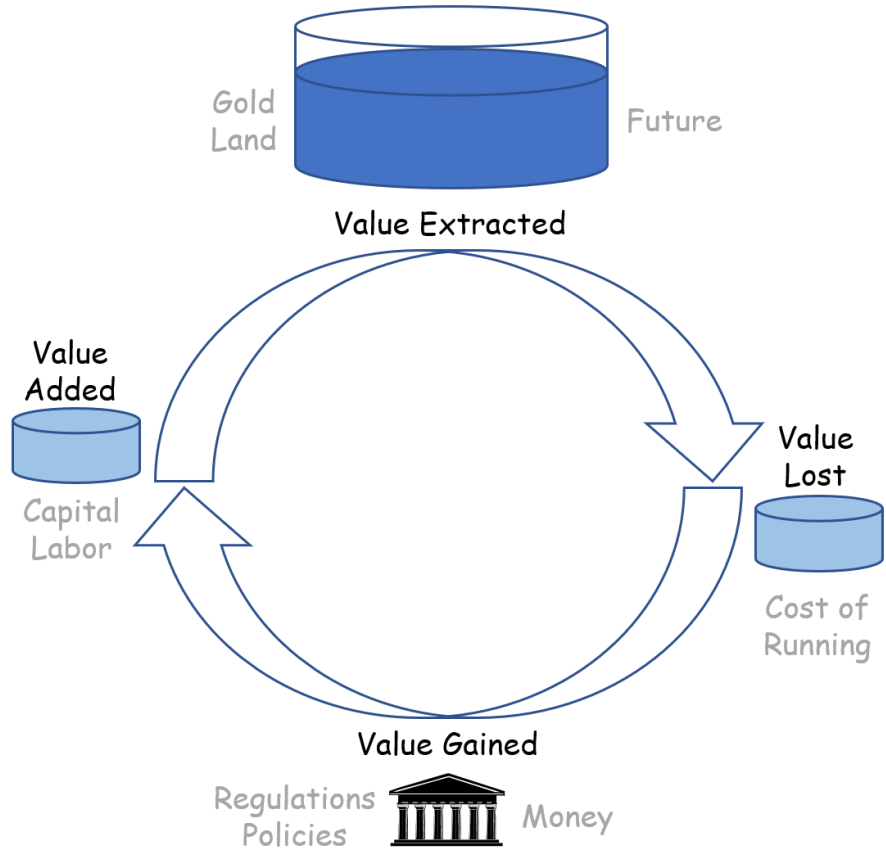

Fig. 11. The government cycle of value

Fig. 12 shows how the money the government prints circulate in the economy. The figure depicts the circle of value for an organization (or individual) that borrows from a bank (red icon). The money is invested along with other capital and labor in the form of added value to extract value from, in this case, natural resources. In the process, some value is lost and the residual value, following the law of conservation of value, becomes the VG for the organization. This value can later be extracted by other organizations or individuals as it becomes accessible or consumed for organizational functions. The bank acquired its VE from the central bank (black icon) by selling securities/debt in exchange for money. The role of the central bank as representative of the (US) government and by extension the people was simply to access future value (a promise really) and convert it into money (VG for the central bank) by adding value through capital and labor required for its operations (mainly the cost of printing the money). To account for its losses, the central bank will receive funds from the government and will gain interest when the banks repay their loans.

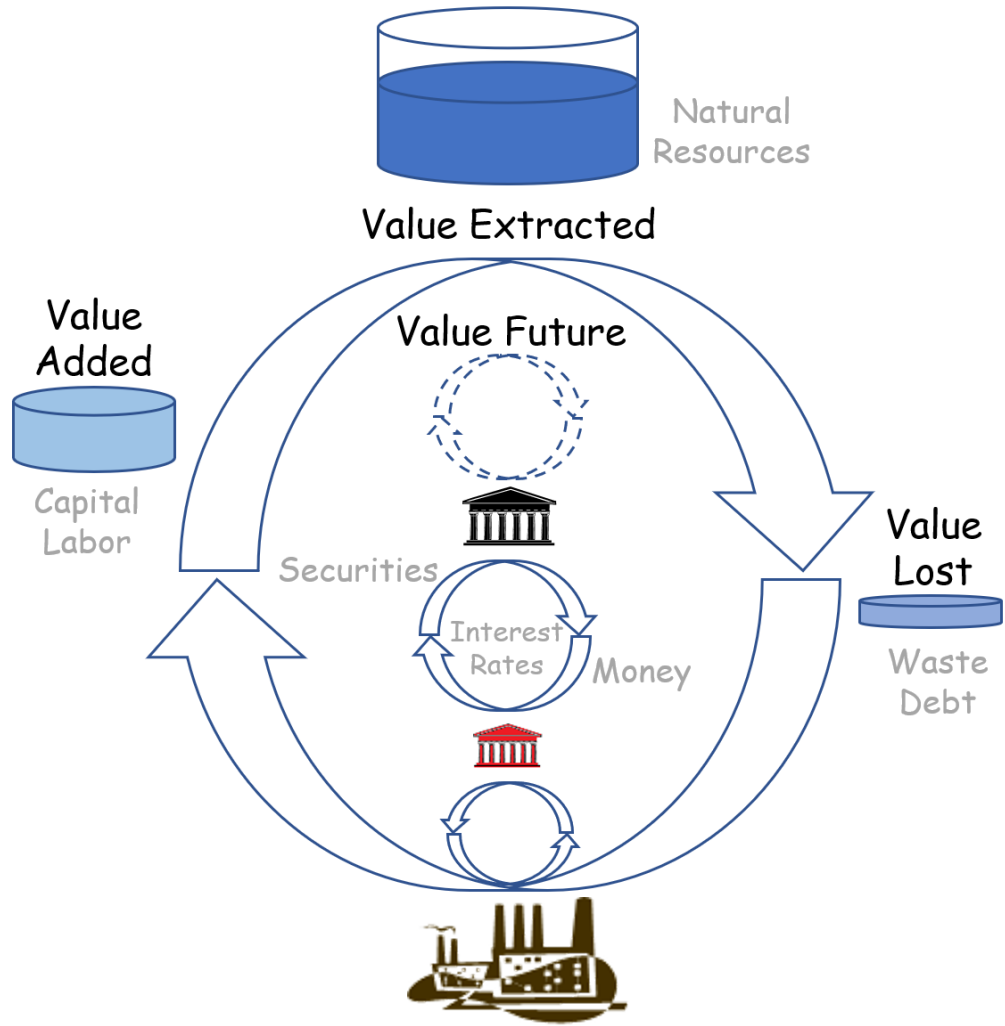

Fig. 12. The flow of value

For the cycle to work properly, uncertainties need to be eliminated or at least addressed in real time. The whole system is based on the trust that all entities will pay back their dues when the time comes. This is not what always happens, especially in cases of economic uncertainties when banks overextend their "trust" lending money that they are not getting back when businesses go bankrupt. As a result, banks can go bust and unable to fulfil their commitment to return the money they took from the central bank/government. The government experiences a shortage in its balance sheet that it typically covers by drawing more from the future (printing more money) or borrowing for other governments and or individuals (like by issuing bonds). The purpose of the government as controller and coordinator is to stabilize prices, moderate long-term interest rates, reduce unemployment, and in general sustain the economy through cycles of value.

A case might help highlight the previous discussion. Let us apply the cycle of value in the case of the government subsidizing an industry like the shale oil and gas industry by making available low interest rates. The tipping balance in Fig. 13 shows the effect of lowering interest rates (red arrow).

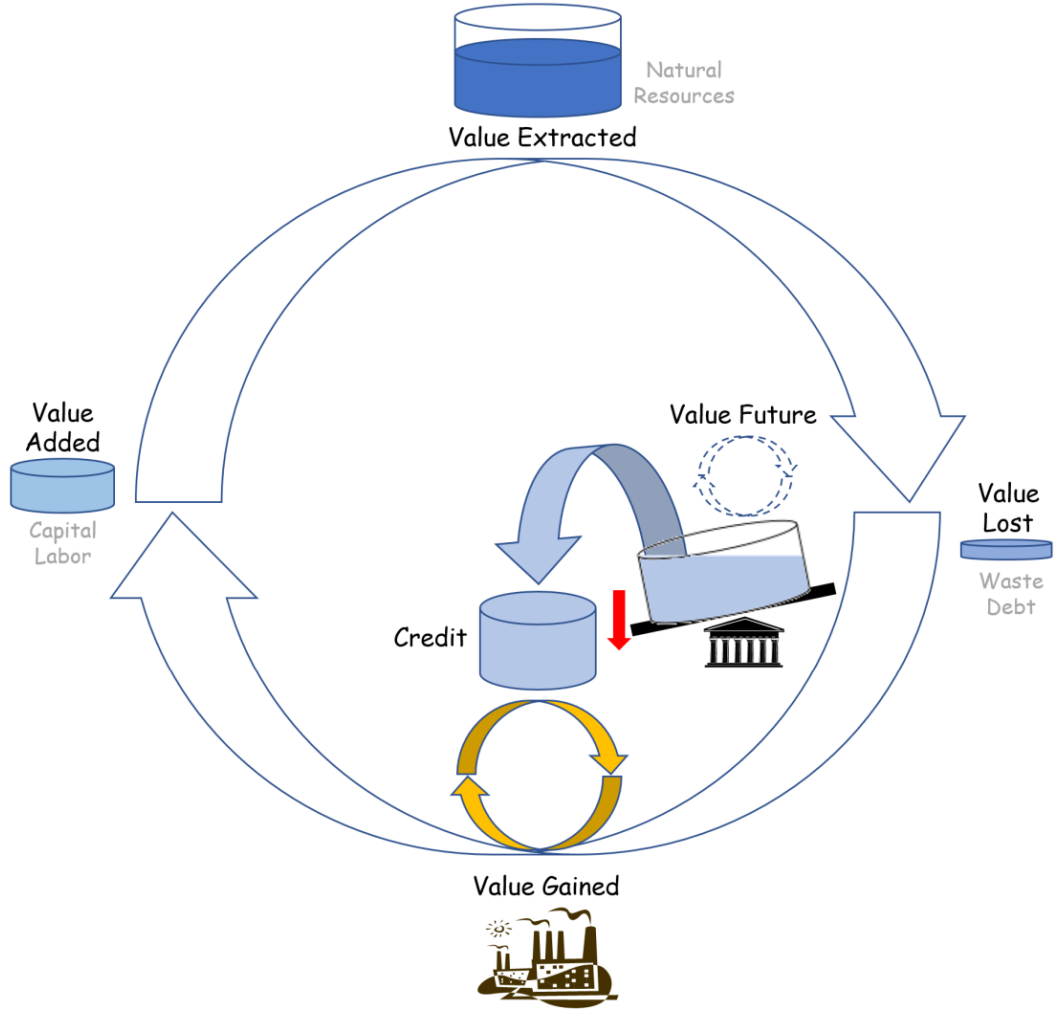

Fig. 13. Cycles of value for shale oil and gas companies

Let us now apply the cycle of value for the organization and the two pools of value it can access, the natural and the government-created credit. For the former (outside cycle in Fig. 14) the index "n" is assigned to indicate the "natural" cycle or the one that accesses natural resources. We have:

$$VA_n + VE_n = VL_n + VG_n$$

For the internal cycle (expanded in Fig. 14) the index "g" is assigned to indicate the government-supported (banking sector) cycle. We have:

$$VA_g + VE_g = VL_g + VG_g$$

By adding the last two equations we get:

$$VA_n + VA_g + VE_n + VE_g = VL_n + VL_g + VG_n + VG_g$$

The goal of the organizations would presumably be to maximize $VG_n + VG_g$. This can be achieved when:

VGn´ + VGg´ = 0

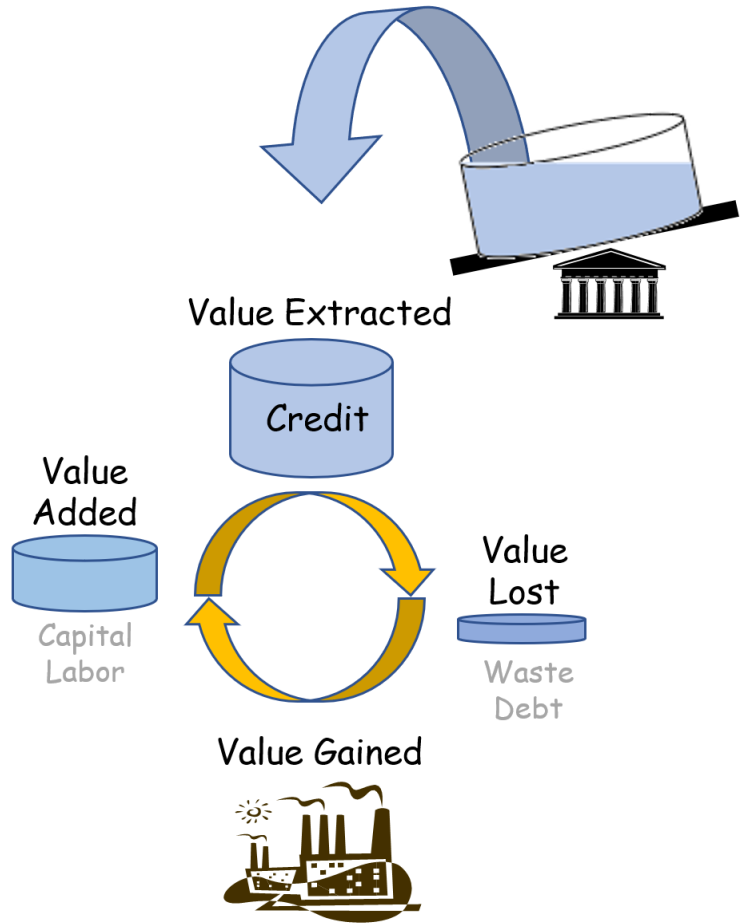

Fig. 14. Government-subsidized cycle of value

It would be relatively easy to see, especially based on historical evidence (van Mourik & Harkiolakis, 2022), that VGg´ is far easier to increase especially when the organization has the right connections that can ensure loans from the banking/government system. From the government point of view the question is when it should pull out the extra pool of value that it made available. The typical answer here is when the industry can sufficiently sustain itself through the natural cycle. Considering the available natural resources as constant or at least accessible at a constant rate (the average availability of wind or sunlight won't change), then for the nature cycle (outside cycle in Fig. 13) we can consider the law of conservation of value in its marginal form:

VAn´ + VEn´ = VLn´ + VGn´

For an abundant nature we can assume:

VEn´ = 0     (7)

Combining the last two equations we get:

$VAn' = VLn' + VGn'$

or

$VGn' = VAn' - VLn'$

For $VGn' > 0$

We get:

$VAn' > VLn'$

This means that gains will keep accumulating as long as the Marginal Value Added is higher than the Marginal Value Lost (all other influencing factors excluded). In the limit the two sides of the inequality will be equal:

$VAn' = VLn'$    (8)

For the government-supported cycle now, the marginal form of the law of conservation of value will take the form:

$VAg' + VEg' = VLg' + VGg'$

or

$VEg' = VLg' + VGg' - VAg'$

If we consider the government's point of view here, then:

$VEg' < 0$

In other words, the government would want to decrease the amount of subsidizing available to the industry. In that case the last two equations give:

$VLg' + VGg' - VAg' < 0$

or

$VLg' + VGg' < VAg'$

Considering the limit when the inequality becomes equality, we get:

$VLg' + VGg' = VAg'$    (9)

Let us consider not the marginal form of the law of conservation of value for both cycles:

$VAn' + VAg' + VEn' + VEg' = VLn' + VLg' + VGn' + VGg'$

Applying (7) we get:

$VAn' + VAg' + VEg' = VLn' + VLg' + VGn' + VGg'$

Applying (8) we get:

$VAg' + VEg' = VLg' + VGn' + VGg'$

Applying (9) we get:

$$VEg' = VGn'$$

In other words, the government should start reducing the subsidies when the rate of the credit it provides is equal to the rate of gain the industry makes from the natural cycle. Past that point the government has no reason to keep supporting an industry. Having said that, if the industry is critical to the survival of its citizens, the government should revert to other means of supporting the industry, including the option of nationalizing it as it does with security and healthcare services.

**9. Government as an Organization**

In this section, the cycle of value from the perspective of the government as an independent/ "selfish" organization will be discussed. The challenge here is the dual role government plays as it is both the recipient and provider of value. While the value creation part has been discussed to some extent in the previous sections, we will see the government here as a consumer of value. As such, the value pool it can access includes the economy and natural resources (Fig. 15). In terms of the economy, the government can access the wealth of its citizens and the organizations within its jurisdiction through various forms of taxation (income taxes, insurance contributions, etc.) and returns on investments (like interests on loans, stocks in businesses, etc.) as well as by accessing the global market like by borrowing from other countries and institutions. In terms of natural resources, this is typically done by its proxies that could include nationalized organizations, institutes (academic, research, etc.), and subsidies. Finally, another resource the government can access is the goodwill and compliance of its citizens.

The government can access the aforementioned pools by adding value in the form of establishing policies, laws, and mechanisms for their enforcement, as well as by providing services citizens need like healthcare, security, education, etc. In the process, the government will lose value in the form of cost for running its operations (public sector) and fulfilling its obligations to its internal and external stakeholders (for example, like paying debt). Eventually, the government will gain value in the form of taxes and income from its subsidies as well as in the form of confidence/votes of its citizens. The role of the ultimate controller, in the case of democratically elected governments, is the voting population.

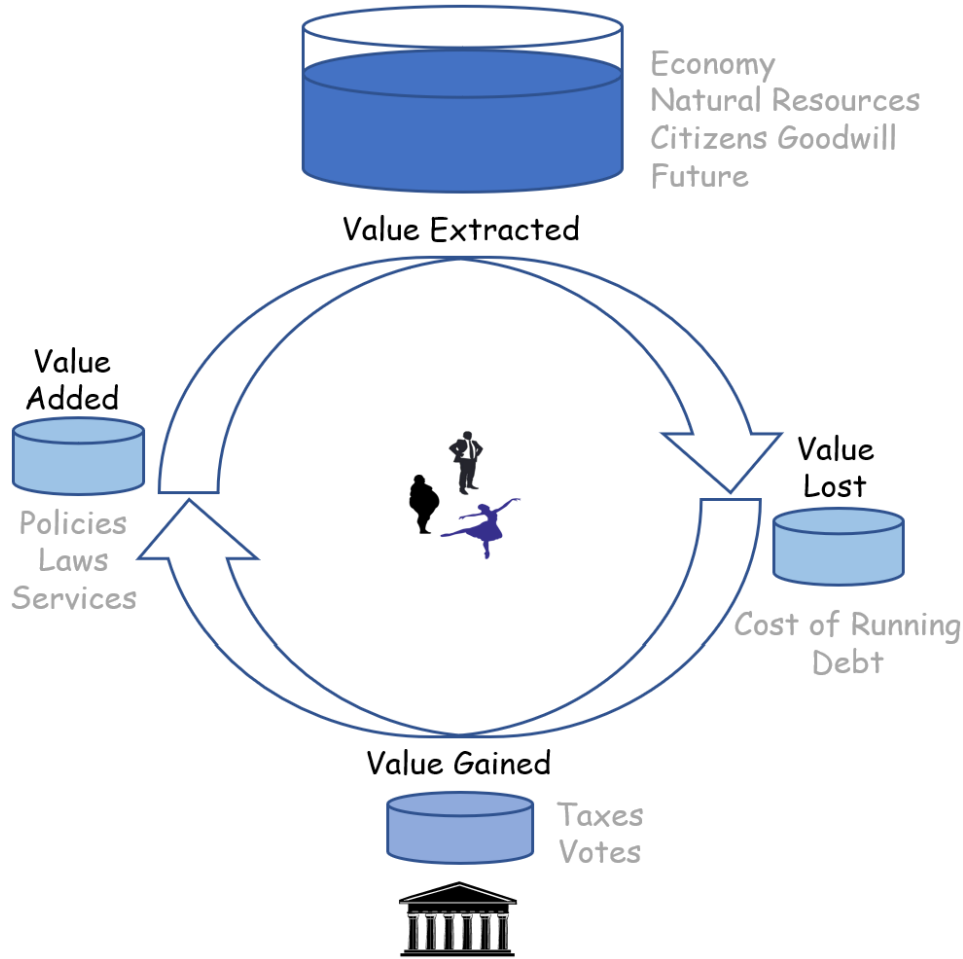

Fig. 15. Government as an organization

While the cycle of value for the government is no different from any other value cycle, it is worth describing the high-level internals of the cycle. One such possibility is presented in Fig. 16 where the cycle is decoupled into two interconnected cycles. When a political party is elected, it forms a government. For simplicity here we will assume the political party got full majority and the government it formed is its proxy. So, the inner cycle starts with the vote of confidence the government received. The government then exerts some effort by adding value (VA) through laws and policies enacted by its various ministries to access the pool of VE. This VE could be in the form of natural resources accessed by nationalized industries or through subsidies to businesses that access the natural resources on its behalf. One such example could be a nuclear power station the government operates while another could be the financial support the government provides to the renewables industry to support a carbon free economy. The VE for the government could also be the economy it controls as it can impose taxes or by selling/rending resources like land, securities, etc. In the process, the government is losing some value mainly due to running expenses, and obligations to its creditors and constituents. The result is VG (summarized as taxes in Fig. 16).

This "inner" part of the value cycles can be expressed as (the lower-case index g indicates government):

$$VA_g + VE_g = VL_g + VG_g$$

In part or in total, the value the government gained up to now is encapsulated for convenience as taxes and goes towards services it provides to its citizens. This can be seen as the VA the government contributes in order to access the goodwill of its citizens (playing the role of VE). Again, some value will be lost, typically as the cost for supporting the public sector. In return, the government gains the confidence of its citizens translated as votes at election time and support in-between (like, say, no strikes).

This "outer" part of the value cycles can be expressed as (the lower-case index c indicates citizens):

$$VA_c + VE_c = VL_c + VG_c$$

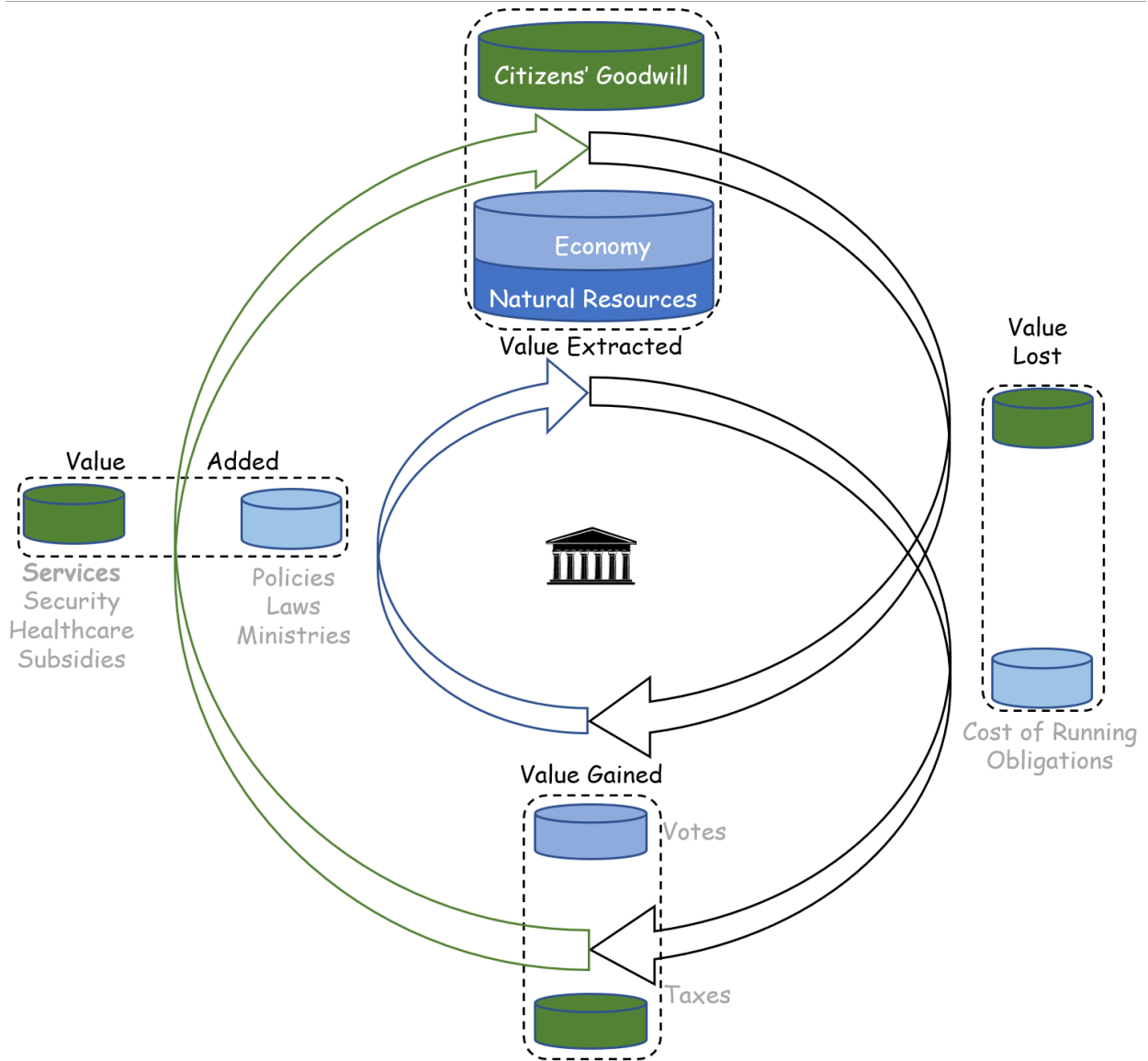

Fig. 16. Interconnected cycles of value

Adding the two cycles we get:

VAg + VAc + VEg + VEc = VLg + VLc + VGg + VGc

Considering marginal components/differentiating the previous equation, we get:

(VAg)´ + (VAc)´ + (VEg)´ + (VEc)´ = (VLg)´ + (VLc)´ + (VGg)´ + (VGc)´ (10)

At this point we can investigate some interesting scenarios. Considering the government as a "selfish" entity, it would want to maximize its VG. Maximizing a quantity means having its first derivative as zero. In this case, this means that:

(VGg)´ + (VGc)´ = 0 => (VGg)´ = - (VGc)´ (11)

This means that the maximum gains for the government are achieved when the rate of change of taxes (margin of taxes) is opposite to the rate of change of votes/citizen confidence (all other factors excluded). The more taxes a government is imposing, the more negative votes it will receive. This is a profound reality that almost anyone will agree with. Nevertheless, it is a natural outcome of the analysis performed here.

Assuming that the government can really control its size/public sector, then for sustainable growth its running cost (VL)´ has to be equal to the rate of value it extracts from the economy plus the confidence it gains from its citizens. This analysis has excluded innovations and technological breakthroughs that could tip the balance of the various elements primarily by decreasing the amount of VA required to produce a certain VG and/or by decreasing the VL. The latter would in effect reduce the cost of running the public sector (public employees' salaries, resources, etc.). In cases where the economy cannot provide as expected, the government can draw value from other governments (getting loans), markets (issuing bonds), and even the future by printing money, unless of course it is willing to reduce its public sector size.

## 10. Forms of Value

Although in the preceding section value was considered as potential to act in an environment and as representing a "physical" entity that can be exchanged between actors/pools of it, traditional economic entities will be considered here as proxies for value to showcase the possibilities of the law of conservation of value (1).

### 10.1. Value as Capital

As a first case in point, value will be equated to capital. For example, consider depositing an X amount of money in a savings account that offers r% interest rate (compounded monthly) while a monthly maintenance fee of Y amount of money is applied. In the context of the cycle of value (assume one calendar year), it can be easily seen that at the end of the cycle/year:

$VA = X$

$VE = rX$

$VL = 12Y$

In order to calculate the gained value, the law of conservation of value (1) can be applied:

$VA + VE = VL + VG$

or

$VG = VA + VE - VL$

or

$VG = X + rX - 12Y$

$VG = X(r+1) - 12Y$

For the second year the previous amount will be our Value Added:

VA = X(r+1) – 12Y

VE = r[X(r+1) – 12Y]

VL = 12Y

so

VG = X(r+1) – 12Y+ r[X(r+1) – 12Y] – 12Y

or

VG = X(r+1) – 12Y + rX(r+1) – r12Y – 12Y

or

VG = X(r+1)(n+1) – 12Y(r + 2)

VG = X(r+1)2 – 12Y(r + 2)

It might be evident that as we keep compounding, the first term will be the classical compounding formula for principle X after a number of years t:

X[(r/12) + 1]12t

while the second term will represent the losses.

### 10.3. Value as price in supply and demand

Considering the supply and demand case, we can assume the price for supply (ps) represents the Value Extracted while the price for demand (pd) represents the Value Gained in a cycle of value. By introducing a linear relationship of price with quantity and kd and ks the proportionality coefficients (rates of change of price with quantity), we can assume:

Demand line: pd = kd * qd + cd

or

VE = kd * qd + cd

Supply line: ps = ks * qs + cs

or

VG = ks * qs + cs

The cycle of value for the transaction will be:

VA + VE = VL + VG

or

VA + kd * qd + cd = VL + ks * qs + cs

If we assume that the losses come primarily from the demand case as taxes and other expenses (we presume whatever products remain will be eventually sold), then VL will be proportional (assume linearly) to quantity by kd. Investment wise, VA will be proportional (assume linearly) to quantity by a function of ks.

In this case the previous equation will become:

$ks * qd + kd * qd + cd = kd * qs + ks * qs + cs$

or

$(ks + kd) * qd + cd = (kd + ks) * qs + cs$

At the equilibrium point we will have VE = VG and qd = qs, so:

VA = VL

or

$ks * qd = kd * qs$

or

ks = kd

This suggests that the rates of production and consumption will be equal at equilibrium.

## Conclusions

An attempt was made in this paper to provide a framework for studying the economy based on the assumption that value as potential to act in an environment is conserved in closed economic systems. This assumption led to the formulation of the law of conservation of value where the sum of value added and value extracted equal the sum of value lost and value gained. As a result of this formulation, typical economic variables like marginal utility and the law of diminishing marginal utility are replaced with marginal value and speed of marginal value in the analysis.

The application of the law of conservation of value to the producer–consumer scenario, as well as the market and economy levels, reveals some already known realities as outcomes of the law of conservation of value and not through intuition. As a result, the use of the formulas developed here can form the basis for further analysis. The recommendation for future research on economic theory would be to ensure that variables chosen to represent value should comply with the law of conservation of value as presented here. This ensures the accounting of the various value terms properly balance out.

## Funding and/or Conflicts of interests/Competing interests.

The author has no relevant financial or non-financial interests to disclose and has no competing interests to declare that are relevant to the content of this article.